\documentclass[sigplan,10pt,nonacm]{acmart}
\usepackage{tikz}
\usepackage{enumitem}
\usepackage{algorithm}
\usepackage{algorithmicx}
\usepackage{algpseudocode}
\usepackage[normalem]{ulem}
\usepackage{filecontents}
\usepackage{xspace}
\usepackage{placeins}
\usepackage{pgfplots}
\usetikzlibrary{matrix, positioning}
\pgfplotsset{compat=1.18}



\newcommand*\circled[1]{%
  \tikz[baseline=(char.base)]{
    \node[shape=circle,draw,inner sep=0.6pt, minimum size=10pt, font=\footnotesize, line width=0.3pt] (char) {#1};}}

\newcommand*\bcircled[1]{%
  \tikz[baseline=(char.base)]{
    \node[shape=circle,fill=black,text=white,inner sep=0.6pt, minimum size=10pt, font=\footnotesize\bfseries] (char) {#1};}}

\newcommand{\sysname}{\texttt{DeltaServe}\xspace}

\definecolor{claudeorange}{RGB}{220,120,30}
\definecolor{claudered}{RGB}{200,40,40}

\begin{document}

\title{DeltaServe: Host-Agnostic Co-Serving of \\
Inference and Fine-Tuning for LLMs}

\author{Jiaxuan Chen}
\affiliation{\institution{McGill University}\country{Canada}}
\author{Jianshu She}
\affiliation{\institution{MBZUAI}\country{United Arab Emirates}}
\author{Ye Yuan}
\affiliation{\institution{McGill University}\country{Canada}}
\author{Rajat Ghosh}
\affiliation{\institution{Nutanix, Inc.}\country{USA}}
\author{Karan Gupta}
\affiliation{\institution{Nutanix, Inc.}\country{USA}}
\author{Qirong Ho}
\affiliation{\institution{MBZUAI}\country{United Arab Emirates}}
\author{Xue Liu}
\affiliation{\institution{McGill University}\country{Canada}}
\author{Oana Balmau}
\affiliation{\institution{McGill University}\country{Canada}}

\begin{abstract}
LLM serving systems are provisioned for peak load to meet strict latency targets, leaving substantial GPU compute idle whenever traffic falls below peak. We present \sysname, a host-agnostic co-serving design that converts this idle inference capacity into LoRA fine-tuning throughput while preserving inference service-level objectives (SLOs). \sysname integrates with existing inference engines through a compact hook interface that requires only multi-LoRA batching support. It exploits the shared execution structure of inference prefill and LoRA fine-tuning forward passes, and uses an SLO-aware scheduler to admit and execute fine-tuning only when sufficient inference headroom is available. The scheduler is driven by a CUDA-graph-aware latency model calibrated offline and refined online. \sysname encapsulates its co-serving mechanisms behind a compact hook interface that requires only multi-LoRA batching support. We integrate \sysname with vLLM, SGLang, and S-LoRA. On a production trace from Nutanix, \sysname on vLLM delivers $2.9\times$ higher fine-tuning throughput than LLMStation at $100\%$ inference SLO compliance, versus $85\%$ for LLMStation. It also achieves $39\%$ higher fine-tuning throughput than a baseline running \emph{vLLM+torchtune}, using no additional hardware and maintaining full SLO compliance.
\end{abstract}

\maketitle

\section{Introduction}

\begin{figure}[t!]
  \centering
  \includegraphics[width=\columnwidth]{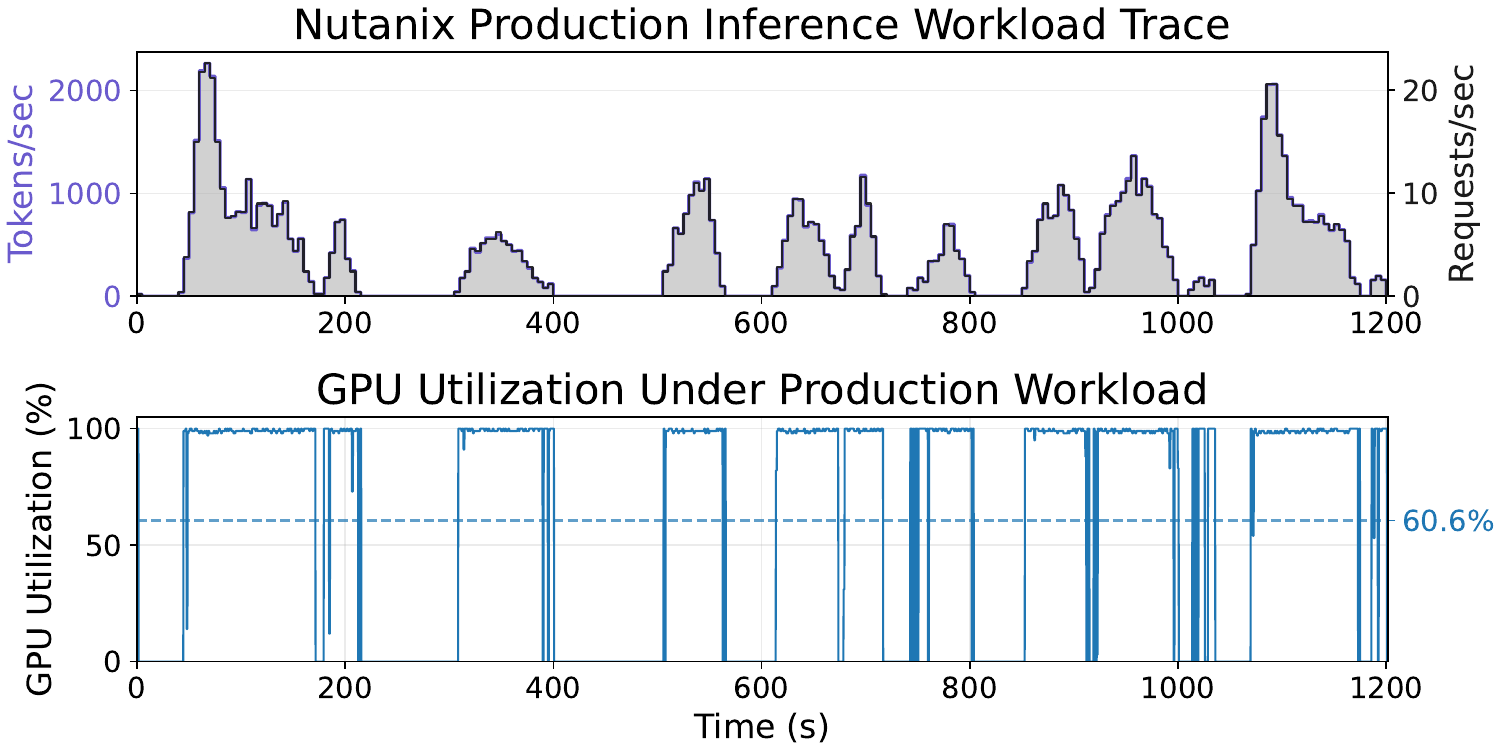}
  \caption{Inference fluctuation in a production workload (Nutanix). On average, 40\% of the GPU is unused.}
  \label{fig:real-world}
\end{figure}

Production LLM serving infrastructure is provisioned for peak inference load so that interactive latency targets can be met during traffic spikes~\cite{288582}. Consequently, outside of those spikes the same hardware sits substantially underutilized. For instance, Figure~\ref{fig:real-world} shows a representative twenty-minute window from the production inference trace of Nutanix, running a co-pilot style coding service. The average load is roughly $60\%$ of provisioned capacity, and the same pattern persists across six months of collected traces. More broadly, the combination of bursty request arrivals~\cite{280922} and the memory-bandwidth-bound decode phase~\cite{10.5555/3618408.3619696, kwon2023vllm} leaves a substantial fraction of GPU compute idle when inference is the only workload running. 

Prior work has approached this underutilization from two angles. The first improves the inference pipeline itself: \textit{Splitwise}~\cite{10.1109/ISCA59077.2024.00019} disaggregates prefill and decode across machines and \textit{Orca}~\cite{280922} mixes prefill and decode requests within the same batch. These techniques raise efficiency under a fixed request stream but do not recover GPU cycles that sit idle when the request stream itself falls below peak. The second co-locates an independent workload on the same GPU using classic sharing primitives such as temporal time-slicing or spatial partitioning (e.g., NVIDIA MIG and MPS). However, these mechanisms operate at granularities far too coarse to track the sub-second fluctuations of inference demand without violating latency targets.

A more promising direction is to use idle inference capacity for a workload that many deployments already need but cannot always provision separately: continual model adaptation. LLM services often need to incorporate new data, evaluate model or adapter variants, and specialize behavior for different products, users, or tenants. For teams with limited GPU budgets, dedicating separate accelerators to this fine-tuning pipeline can directly reduce the capacity available for serving, forcing a trade-off between keeping models up to date and meeting inference latency targets.

Parameter-efficient fine-tuning (PEFT)~\cite{peft}, and Low-Rank Adaptation (LoRA)~\cite{hu2022lora} fine-tuning in particular, is well matched to this setting because it shares most of its execution state with inference. With LoRA, the base model remains frozen and only a small adapter is updated, so inference and fine-tuning share frozen weights, kernels, and memory. The forward pass of a LoRA fine-tuning step is structurally the same as inference prefill, except that fine-tuning must preserve activations for the later backward pass~\cite{10.1145/3458817.3476209}. Two recent systems explore this opportunity. LLMStation~\cite{he2025resource} co-executes PEFT forward work with inference decode iterations to exploit decode's memory-bandwidth-bound compute slack. FlexLLM~\cite{oliaro2026flexllm} interleaves the two workloads at the level of token chunks, slicing fine-tuning sequences and running each chunk alongside inference tokens. Both demonstrate that LoRA co-serving is feasible while preserving inference SLOs, but each leaves room for improvement: LLMStation's decode fusion is constrained by tight per-step time-per-output-token (TPOT) budgets and a latency cache that does not distinguish CUDA-graph from eager execution, while FlexLLM's chunked training pays substantial per-chunk data-movement overhead.

Exploiting this opportunity well requires fine-grained control across scheduling, execution, and memory management. The first challenge is that fine-tuning forward and backward computation consume the same compute and memory bandwidth that inference relies on, so uncontrolled admission inflates time-to-first-token (TTFT) for new requests and time-per-output-token (TPOT) for ongoing decodes~\cite{280922, 10.5555/3618408.3619696, kwon2023vllm}. Second, backward computation introduces long-running GPU work that must be interleaved with inference without disrupting the kernel-launch and execution efficiency of the latency-critical serving pipeline. Third, inference load fluctuates at sub-second timescales, so fine-tuning must be admitted, throttled, paused, and resumed at a finer granularity than existing co-serving systems support. 

A fourth challenge is practical rather than algorithmic: the technique should be easy to add to an existing inference engine, not delivered as a new serving system that replaces it. Production inference already runs on mature, heavily optimized engines, and these engines differ widely in how they schedule and execute requests. A co-serving technique that is tied to one engine's internals, or that requires substantial re-engineering to adopt, is unlikely to be used in practice. 

We present \sysname, a host-agnostic co-serving design that augments an existing inference engine with LoRA fine-tuning under explicit SLO control. 
\sysname is not a standalone serving system but a design that can be incorporated into an existing engine. Its co-serving logic, comprising the SLO latency model, the admission policy, and the decoupled backward executor, is host-agnostic and is exposed through a compact interface of integration hooks. The only capability \sysname requires of the host engine is multi-LoRA batching, which modern serving systems already support in order to serve many adapters over a single frozen base model. We integrate \sysname into three such engines spanning the range of serving stacks it might attach to: vLLM and SGLang, two state-of-the-art inference engines widely deployed in industry, and S-LoRA, a popular research serving system built around multi-LoRA batching, the one capability \sysname requires. Each integration reuses the same host-agnostic core and rewrites only the engine-specific hooks.

\sysname exploits the structural identity between inference prefill and the LoRA fine-tuning forward pass: the two perform the same computation, so fine-tuning folds into ongoing inference work rather than running as an unrelated job.
At the core of \sysname is an SLO-aware admission scheduler built on this observation, driven by an analytical model that predicts how much latency the fine-tuning would add to the inference workload \textit{at the batch-granularity} before it is admitted. This prediction gives \sysname the fine-grained control prior co-serving systems lack: it admits, throttles, and pauses fine-tuning as load shifts, always within the latency budget of in-flight and queued requests, thereby preserving TTFT and TPOT targets. The backward pass runs in a separate GPU process that gives way to inference whenever inference demands more compute, to let fine-tuning consume only otherwise-idle capacity.

\vspace{1mm}
\noindent In summary, the contributions of this paper are:
\begin{itemize}[leftmargin=1em, itemsep=0.1em, topsep=2pt, partopsep=0pt]
\item \textbf{A host-agnostic co-serving design.} \sysname decomposes SLO-aware co-serving of inference and fine-tuning into a reusable, host-agnostic core and a compact set of integration hooks, requiring only multi-LoRA batching from the host engine; we showcase the same core on vLLM~\cite{kwon2023vllm}, SGLang~\cite{zheng2024sglang}, and S-LoRA~\cite{MLSYS2024_906419cd}.

\item \textbf{Design and implementation of \sysname}, consisting of the following three novel technical contributions:

\textbf{(1) SLO-aware fine-tuning admission.} Building on the structural identity between inference prefill and the LoRA forward pass, \sysname incorporates fine-tuning into ongoing inference batches and admits it against per-request TTFT and TPOT budgets, preserving inference SLOs.

\textbf{(2) A CUDA-graph-aware latency model.} A light analytical model, calibrated offline and refined online, predicts per-step execution time under both graph and eager execution and provides the basis for admission decisions.

\textbf{(3) A decoupled backward executor.} Fine-tuning backpropagation executes in a separate GPU subprocess that yields to inference at each transformer-layer boundary and is preempted upon the arrival of new requests, consuming only otherwise-idle compute.

\item \textbf{Evaluation.} We implement \sysname on three serving engines, vLLM, SGLang, and S-LoRA, confirming that the same host-agnostic core ports across them, and we release it as open source. On a production trace from Nutanix, \sysname-vLLM sustains $2.9\times$ the fine-tuning throughput of LLMStation, the state-of-the-art co-serving system, and $39.84\%$ more than a split-pool baseline that dedicates a separate GPU to fine-tuning, all at full SLO compliance and without additional hardware.
\end{itemize}

\section{Background and Related Work}
\label{background-v2}

\subsection{ML Serving and Inference Systems}
\label{background-inf-peft-v2}
LLM inference is autoregressive: output tokens are generated one at a time, each conditioned on the prompt and all previously generated tokens. Serving engines advance generation in discrete \emph{steps}, or iterations, where each step runs one forward pass over the current batch and emits one token per active request.  Each request proceeds through two phases with distinct resource profiles~\cite{kwon2023vllm, 10.1109/ISCA59077.2024.00019}. In \textit{prefill}, the engine processes the full prompt to produce the first output token and initialize the key–value (KV) cache. This full-sequence computation is primarily compute-bound~\cite{dao2023flashattention2fasterattentionbetter}. In \textit{decode}, each step processes only the newest token but reads the accumulated KV cache, making it lower in arithmetic intensity and often memory-bandwidth-bound~\cite{10.5555/3618408.3619696}. These phases correspond to the main latency objectives: time-to-first-token (TTFT), largely driven by prefill, and time-per-output-token (TPOT), determined by repeated decode steps.

Inference serving engines are part of a large body of systems research targeting efficient ML serving at scale. Earlier frameworks~\cite{crankshaw2017clipper, 258862, 276950, 222629, 273804} address latency modeling, dynamic scaling, memory swapping~\cite{258864}, preemptive scheduling~\cite{9065590}, workload prediction~\cite{285173}, and cost-oriented provisioning~\cite{234998}, while Nexus~\cite{shen2019nexus} and Gpulet~\cite{280768} add finer-grained batching and GPU virtualization for small-model deployments. More recent LLM serving stacks, including vLLM~\cite{kwon2023vllm}, Sarathi~\cite{agrawal2023sarathiefficientllminference}, SGLang~\cite{zheng2024sglang}, AlpaServe~\cite{288582}, and Punica~\cite{MLSYS2024_054de805}, optimize decoder-heavy traffic through continuous batching, KV-cache management, and customized PEFT kernels, and others add fairness-aware~\cite{kumar2025aqua} or speculative-decoding~\cite{chen2023acceleratinglargelanguagemodel, miao2024specinfer, oliaro2025suffixdecoding, li2025adaserveacceleratingmultislollm, liu2025turbospecclosedloopspeculationcontrol} scheduling. \sysname does not compete with these stacks but builds on them, treating engines such as vLLM and SGLang as hosts, reusing their inference pipeline, and adding only the machinery to co-serve fine-tuning.

\subsection{Parameter-Efficient Fine-Tuning}
Full-parameter fine-tuning of large language models is expensive because gradients and optimizer states must be maintained for billions of parameters. Parameter-efficient fine-tuning (PEFT)~\cite{peft} reduces this cost by freezing the base model and training only small additional modules.

Low-Rank Adaptation (LoRA)~\cite{hu2022lora} is a widely used PEFT method that injects trainable low-rank updates into selected layers of the base model, typically the attention and feed-forward projections. Given a frozen weight matrix $W \in \mathbb{R}^{d \times k}$, LoRA represents the trainable update as $\Delta W = BA$, where $A \in \mathbb{R}^{r \times k}$ and $B \in \mathbb{R}^{d \times r}$ with rank $r \ll \min(d,k)$. The effective weight becomes $W_{\text{eff}} = W + BA$, while only the low-rank factors $A$ and $B$ are updated during fine-tuning. Since these factors are much smaller than $W$, LoRA substantially reduces the memory required for gradients and optimizer states compared with full-parameter fine-tuning.

PEFT also changes model serving: instead of deploying one full model per task, a serving system can host many lightweight adapters over a shared frozen base model. A naive implementation runs one forward pass per adapter, preventing efficient batching. Punica~\cite{MLSYS2024_054de805} and S-LoRA~\cite{MLSYS2024_906419cd} address this with \emph{multi-LoRA batching}, where tokens using different adapters share one base-model forward pass while applying adapter-specific low-rank weights. This makes multi-adapter LoRA serving a natural substrate for co-serving: fine-tuning samples can be represented as adapter-bound requests and processed through the same adapter-aware batching and forward-execution pipeline as inference.

\subsection{GPU Resource Sharing for ML}
\label{background-gpu-sharing-v2}
Efficient sharing of GPU resources across heterogeneous ML workloads has also been widely studied. Systems such as Lyra~\cite{li2023lyra} focus on dynamically resizing GPU allocations to match workload behavior, whereas Gandiva~\cite{222611} and AntMan~\cite{258957} employ time-slicing to improve cluster-level utilization. GSLICE~\cite{dhakal2020gslice} introduces adaptive partitioning mechanisms to accommodate different latency SLOs. While these approaches substantially improve multiplexing for classic training and inference pipelines, they do not directly address the unique characteristics of PEFT-based LLM services, where numerous lightweight models share a large frozen backbone, nor do they fully resolve the resource inefficiencies observed in modern multi-adapter serving environments. 

\subsection{Co-Serving Systems}
\label{background-coserving-v2}
Co-serving systems for LLMs execute latency-sensitive inference and fine-tuning concurrently on the same GPU resources. For LoRA-based LLMs, this is attractive because both workloads share the frozen base model, while fine-tuning updates only lightweight adapter parameters. However, supporting fine-tuning inside an inference serving system requires more than admitting extra GPU work. It introduces training data queues, activation storage, backward execution, optimizer updates, and adapter-weight synchronization, requiring scheduling and execution structures absent from inference-only pipelines.


LLMStation~\cite{he2025resource} is the closest concurrent co-serving system to \sysname, differing in both scope and mechanism. In scope, it is a serving system built on vLLM, whereas \sysname is a host-agnostic co-serving design that reuses the host's existing serving pipeline as much as possible and adds fine-tuning through a compact set of hooks, so it adapts to mainstream serving engines; we realize it on vLLM, SGLang, and S-LoRA. In mechanism, both co-serve LoRA fine-tuning under inference SLOs, but differ in how fine-tuning is integrated and how its latency is predicted.

LLMStation confines fine-tuning to decode-phase headroom, where each admission must fit decode's tight per-token budget. \sysname instead admits fine-tuning in any phase, modeling each sample as a single-step, prefill-only request (Section~\ref{sec:design}), so it can admit against the full inference latency budget. The two also predict latency differently: LLMStation relies on a per-shape latency cache that ignores the graph-versus-eager distinction, whereas \sysname uses an execution-mode-aware analytical model that needs no cache and prices that distinction directly.

FlexLLM~\cite{oliaro2026flexllm}
is a system that interleaves inference and fine-tuning at token granularity~\cite{pmlr-v139-li21y}, slicing fine-tuning sequences into chunks and running each chunk's forward and backward computation alongside inference tokens. This fine-grained overlap also incurs substantial data movement: each chunk reloads the full model from GPU memory, and when a chunk is left unsplit to avoid that cost, an in-flight fine-tuning step can occupy the pipeline and stall the TTFT of newly arrived requests. \sysname takes a different position, retaining full-sequence fine-tuning forward passes and treating them as prefill-like work rather than decomposing them across tokens; it thus operates at a coarser but more latency-stable granularity that keeps inference transparent while still harvesting idle GPU capacity.

These systems show that LoRA fine-tuning can be colocated with online inference, but also expose a practical limitation: co-serving often requires substantial changes to the serving pipeline. \sysname targets this integration problem by reusing the existing LoRA serving path for fine-tuning forward computation, deferring backward execution outside the critical inference path, and using an accurate SLO-aware scheduler that is independent of a particular host-engine structure. It admits fine-tuning only when the predicted impact remains within TTFT and TPOT budgets. More broadly, unlike LLMStation and FlexLLM, which define their own co-serving execution structures, \sysname attaches to existing engines through compact hooks while preserving the host scheduler and optimized execution path.

\subsection{CUDA Graphs}
\label{background-cuda-graphs-v2}
A CUDA graph records a sequence of GPU kernel launches and replays it with a single host-side submission. This removes per-kernel launch overhead, which is significant for short or memory-bound kernels and especially important for LLM decode, where each step performs limited computation but invokes many kernels. Modern LLM serving systems such as vLLM~\cite{kwon2023vllm} and SGLang~\cite{zheng2024sglang} therefore use CUDA graphs to accelerate inference. Decode can often be captured as a full graph because its per-step shape is relatively stable. Prefill is more difficult because prompt lengths vary, so these systems use piecewise graph execution: shape-stable parts of the forward pass are captured and replayed, while variable-length attention runs eagerly. 

\section{System Design}
\label{sec:design}
The background discussion shows that LoRA co-serving is appealing because inference and fine-tuning share the frozen base model, but realizing it inside an existing serving engine is difficult: fine-tuning needs training-specific control, activation capture, backward execution, and optimizer updates that inference engines are not designed to provide. \sysname addresses this gap by separating what must be added for fine-tuning from what should remain under the host system’s control. Its design keeps the host responsible for request handling, batching, KV-cache management, sampling, and optimized forward execution, while \sysname adds only the mechanisms needed to turn admitted fine-tuning samples into prefill-like batch entries and to process their backward pass outside the critical inference path.

This section describes how \sysname implements this separation. Section~\ref{sec:overview} first presents the host interface and the add-on components that attach to a generic inference engine. The following sections then describe how fine-tuning samples are executed through the host forward path, how the scheduler estimates SLO headroom and admits fine-tuning work, and how the backward subprocess performs training updates while yielding to latency-critical inference.

\subsection{\sysname Overview}
\label{sec:overview}
Figure~\ref{fig:overview} illustrates \sysname as an extension around a generic LLM inference engine. The white boxes represent the common structure of modern serving systems: a client-facing request queue, a scheduler, and a GPU execution engine that executes the model forward pass. We use this abstraction because these components appear across serving engines despite their internal implementations differ. \sysname therefore defines its interface at the boundary of these components rather than depending on a particular runtime.

Under this interface, \sysname does not replace the serving pipeline. The host remains responsible for request handling, scheduling, and GPU execution, and \sysname reuses the engine’s optimized serving path, including KV-cache management, inference kernels, batching logic, sampling, and inference optimizations such as CUDA-graph replay when available. The only semantic capability \sysname requires from the host is multi-LoRA batching, which allows a reserved training adapter to share the same forward path as inference adapters. The blue components in Figure~\ref{fig:overview} mark the \sysname components that extend the host pipeline:

\begin{figure}[h!]
  \centering
  \includegraphics[width=\columnwidth]{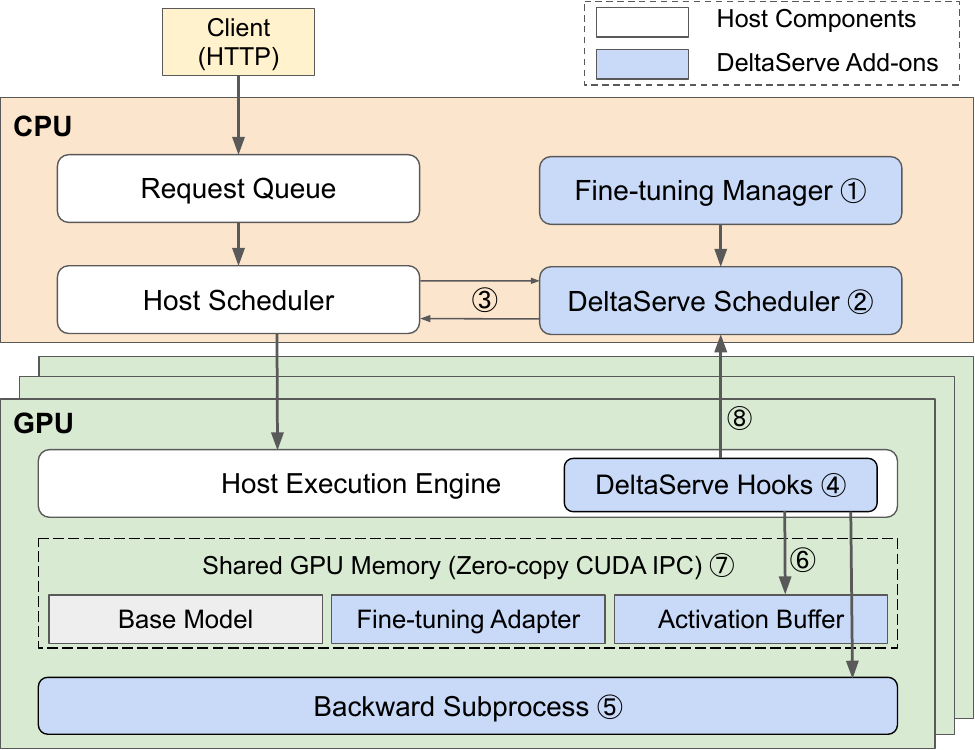}
  \caption{\sysname architecture and host interface. White boxes are generic inference serving system components; blue boxes are \sysname add-ons. Numbered markers indicate the attachment points and their relation described in Section~\ref{sec:design}: CPU-side admission control, GPU-side activation capture, shared GPU memory, and the backward subprocess.}
  \label{fig:overview}
\end{figure}

\begin{figure*}[t!]
  \centering
  \includegraphics[width=\linewidth]{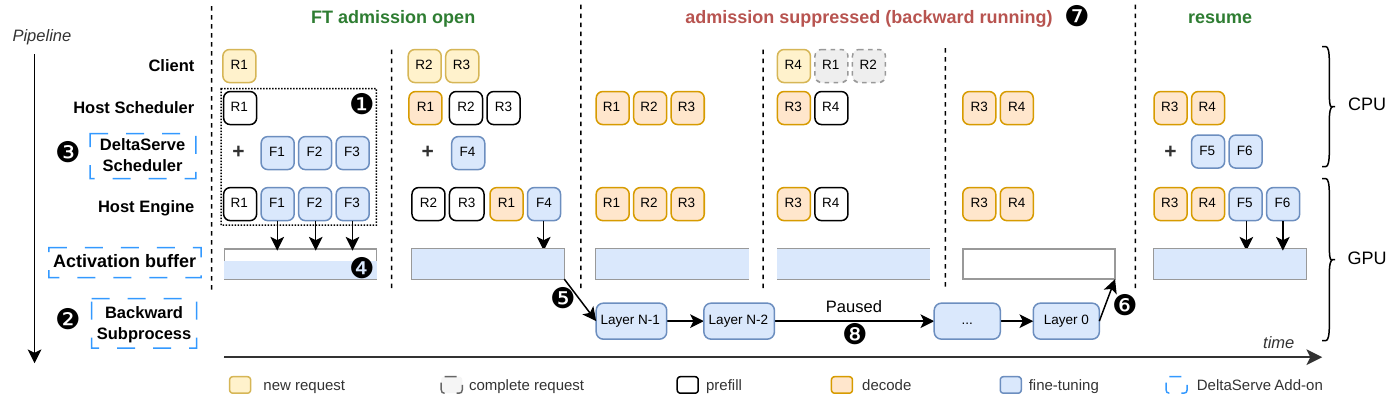}
\caption{Co-serving workflow in \sysname. The vertical axis shows the serving pipeline, and time advances from left to right. Each box represents one request or fine-tuning sample: yellow boxes are new arrivals, white boxes are inference prefill, orange boxes are inference decode, dashed gray boxes are completed requests, and blue boxes are fine-tuning samples. Dashed blue outlines mark \sysname add-on components.}
  \label{fig:pipeline}
\end{figure*}

\begin{itemize}[leftmargin=1em, itemsep=0.1em]
\item \textbf{Fine-tuning Manager} ~\circled{1}.
The fine-tuning manager is an add-on module that supplies the training side of co-serving outside the host engine's inference request queue. It loads and tokenizes the fine-tuning corpus, orders candidate samples for admission, and tracks training progress across epochs. This separation lets \sysname introduce fine-tuning work into the serving loop without modifying the host's client-facing request path.

\item \textbf{\sysname Scheduler}~\circled{2}.
The \sysname scheduler is an extension that intercepts the host's scheduling loop before the host scheduler forms a new batch and dispatches it to its execution engine. It inspects the host's proposed batch, estimates its execution cost with an analytic latency model, and admits as many fine-tuning samples as the inference request's SLO permits \circled{3}. The host then builds a single batch combining both inference and fine-tuning. The admitted fine-tuning samples are retired immediately after a single forward pass, ensuring that fine-tuning outputs never enter the client-facing serving path.

\item \textbf{Inference-Engine Hooks}~\circled{4}.
The inference-engine hooks are lightweight insertions into the host's forward path and engine loop. Model-level forward hooks~\circled{6} save the activations needed for fine-tuning backpropagation into the shared GPU memory~\circled{7}. They also record per-step execution time to refine the latency model online~\circled{9}. The same hook interface also supports an offline profiling pass that initializes the model before serving begins.

\item \textbf{Backward Subprocess~}\circled{5}.
The backward subprocess is an add-on process launched alongside the inference engine under a CUDA MPS control daemon. It supplies the training-specific computation absent from the inference engine, such as gradient computation and adapter weight updates; its mechanics are detailed in Section~\ref{fine-tuning}.
\end{itemize}

\subsection{Co-Serving Model}
\label{fine-tuning}
Figure~\ref{fig:pipeline} illustrates an example workflow by which \sysname admits and executes fine-tuning samples during co-serving. \sysname folds the LoRA fine-tuning forward into the host engine's existing forward pass \bcircled{1} and adds a separate GPU subprocess for the fine-tuning backward \bcircled{2}. The fine-tuning forward shares the host's batch rather than running as an independent pass: as established earlier, the forward pass of LoRA fine-tuning is structurally identical to inference prefill, so the cost of folding a fine-tuning sample into an ongoing inference batch is close to that of one additional prefill sample rather than that of a separate forward pass.  \sysname scheduler \bcircled{3} exploits this property in its latency model (Section~\ref{sec:slo}), which directly prices the added cost and uses it to guide SLO-aware admission.

\vspace{1em}
\noindent\textbf{Mixed forward batch.}
\sysname admits each fine-tuning sample as an ordinary host-batch entry: a single-step, prefill-only request routed to a reserved fine-tuning adapter and configured to produce no client-visible output. The sample executes through the host’s existing multi-LoRA forward path, using the same pipeline as inference requests. \sysname’s only addition to this path is activation capture \bcircled{4}. Hooks on the host model record residual-stream activations for fine-tuning rows into a buffer shared with the backward subprocess, while loss and gradient computation are left entirely to the backward path.

The amount of activation state to retain is a design choice that trades GPU memory for backward recomputation. By default, \sysname saves only the residual-stream input to each transformer layer, since each layer can be recomputed from this input during backward. For a model with $N_L$ layers, hidden size $d$, and activation precision $p$, this costs roughly $N_Ldp$ bytes per fine-tuning token; for a typical 8B model with $N_L=32$, $d=4096$, and fp16 activations, this is about $0.25$\,MB per token. If GPU memory permits, one may cache additional intra-layer intermediates, such as attention projections and feed-forward activations, to reduce recomputation. If memory is constrained, \sysname can offload captured activations to host memory and stream them back on demand. After the step completes, \sysname retires the fine-tuning request before any output token is emitted, keeping fine-tuning invisible to inference clients.

\vspace{1em}
\noindent\textbf{Backward subprocess.}
The fine-tuning backward pass runs in a dedicated GPU subprocess under CUDA MPS, separate from the host forward engine. We use a subprocess rather than an additional CUDA stream so that inference and backward execution maintain distinct CUDA contexts~\cite{10.1145/3627703.3629578}, reducing launch-path contention and interference with latency-critical inference while allowing backward kernels to execute concurrently with inference.

At startup, the subprocess maps the base model, the fine-tuning adapter, and the activation buffers into its address space using inter-process shared GPU memory. It can therefore access all tensors required for training without inter-process copies. The subprocess then waits in a blocking event loop until the scheduler issues a backward task. For each task, it reads the shared activations, reconstructs logits from the saved hidden states \bcircled{5}, recomputing unsaved intermediates, computes the loss and LoRA gradients, and applies optimizer updates directly to the shared adapter weights. After the update completes, it releases the corresponding activation-buffer entries and returns to the idle state \bcircled{6}. While a backward task is in flight, the scheduler admits no new fine-tuning samples, ensuring that activations are not overwritten before consumption \bcircled{7}. Once the task completes, the scheduler records the trained samples and reopens fine-tuning admission.

\subsection{SLO Budgeting}
\label{sec:slo}
A co-serving scheduler must determine, before dispatching a batch, how much inference latency headroom remains. \sysname formulates this headroom as a per-batch \emph{budget}: the difference between the batch’s predicted execution time and the latency deadline imposed by either the earliest-arriving request’s TTFT constraint or the TPOT constraint of ongoing decode requests. \sysname's SLO estimator supplies the prediction, and the scheduler (Section~\ref{section:scheduling}) consumes the budget to decide how much fine-tuning work can be safely co-scheduled with inference.

\vspace{1em}
\noindent\textbf{Analytical latency model.} At deployment time, the model architecture is fixed, so the latency of a forward step is determined primarily by the batch composition. The dominant compute comes from the transformer attention and feed-forward layers: attention cost depends on the sequence lengths processed in the step, while feed-forward cost scales with the number of tokens passing through the layers. In addition to computation, step latency includes memory traffic from key–value cache reads for decode requests and activation-buffer writes for admitted fine-tuning tokens. Thus, for a mixed batch, \sysname estimates step latency from the prompt lengths ${n_i}$ of prefill requests, the number of active decode requests $B_d$, the total key–value cache size $K$, and the number of admitted fine-tuning tokens $T_{\mathrm{ft}}$. \sysname predicts the step time from them as

\begin{equation}
T \approx \alpha \!\sum_{i} (n_i + B_d)^2 + \beta\,( T_{\mathrm{in}} + B_d ) + \gamma\, T_{\mathrm{ft}} + \varepsilon\, K + c ,
\tag{\textit{I}}\label{eq:prefill}
\end{equation}
where $T_{\mathrm{in}} = \sum_i n_i$ is the total prefill length. Here $\alpha$ prices the self-attention compute, whose quadratic span combines each prefill request's $n_i$ prompt tokens with the $B_d$ decode queries sharing the step; $\beta$ prices the feed-forward compute; $\gamma$ captures the cost of saving the fine-tuning activations; $\varepsilon$ captures key–value cache memory access for decode requests; and $c$ is a fixed per-step overhead.

For a step that carries only decode work, the prefill and fine-tuning terms vanish, leaving the feed-forward cost of the $B_d$ decode tokens and the key--value cache access:
\begin{equation}
T \approx \beta'\, B_d + \varepsilon'\, K + c' .
\tag{\textit{II}}\label{eq:decode}
\end{equation}
\noindent The two models are fit independently, with the coefficients of each seeded by an offline profiling pass and refined online during serving (Section~\ref{sec:calibration}). This split matches hosts that form a \emph{prefill-decode mixed batch}, a single step that combines prefill and decode requests and is priced by Equation~\ref{eq:prefill}; it does not require fusing, however: a host that runs prefill and decode as separate steps uses Equation~\ref{eq:prefill} for its prefill steps and Equation~\ref{eq:decode} for its decode steps, so the model transfers across host engines unchanged.

\vspace{1em}
\noindent\textbf{Execution-mode coefficients.}
As discussed in Section~\ref{background-cuda-graphs-v2}, production inference systems often capture CUDA graphs to remove per-kernel launch overhead. One forward pass step can therefore run in one of two modes, replayed from such a captured graph or \emph{eagerly}, with its kernels dispatched one at a time and no graph to replay. Which mode a step uses changes its latency by nearly an order of magnitude at small batch sizes, so the model for a prefill-carrying step must account for both. An inference-only prefill or mixed step can be replayed from the host's captured graph, whereas a co-serving step, which carries fine-tuning tokens, is forced eager because the activation-capture hooks cannot run inside a replayed graph without major modification to the host engine. \sysname therefore fits Equation~\ref{eq:prefill} with two sets of coefficients, one for graph execution and one for eager, and selects between them at prediction time using the host's own graph-versus-eager decision.

\vspace{1em}
\noindent\textbf{TTFT budget.}
With the latency model in place, \sysname determines how much fine-tuning can be admitted without violating latency objectives. For the upcoming step, it inspects the earliest-arriving inference request and computes its remaining TTFT budget,
\begin{equation}
\Delta_{\mathrm{budget}} = T_{\mathrm{SLO}} - (t_{\mathrm{now}} - t_{\mathrm{arrival}}),
\tag{\textit{III}}\label{eq:budget}
\end{equation}
\noindent where $T_{\mathrm{SLO}}$ is the request's TTFT target. This conversion of an SLO target into a dynamic per-step budget is the central admission lever; Section~\ref{section:scheduling} describes how the scheduler turns it into a fine-tuning token count.

\subsection{\sysname Scheduler}
\label{section:scheduling}
Guided by the SLO estimator, the \sysname scheduler serves as the control point between the host engine and \sysname’s fine-tuning components. It operates inside the host’s scheduling loop, intercepting each step before dispatch. At this point, it draws candidate samples from the fine-tuning manager (Section~\ref{fine-tuning}), evaluates the latency impact of augmenting the pending inference batch, and admits fine-tuning only when the SLO budget and activation-buffer capacity permit. The scheduler also coordinates with the backward subprocess: it suppresses new fine-tuning admission while a backward batch is running and updates fine-tuning progress as completed backward batches are reported. 

At a high level, \sysname does not construct a separate fine-tuning batch. Instead, the scheduler enqueues admitted samples into the host’s scheduling path so that the host forms a single combined batch. After the step completes, \sysname retires the fine-tuning entries before results are returned to clients, leaving the host’s inference semantics unchanged. The rest of this section describes how the scheduler determines the amount of fine-tuning work to admit at each opportunity.

\vspace{1em}
\noindent\textbf{Fine-tuning admission.}
Algorithm~\ref{alg:admit} shows the fine-tuning admission procedure for one scheduling step. The input is the inference batch $B_{\mathrm{inf}}$ that the host is about to dispatch, together with the SLO estimator, fine-tuning pool, backward-process state, remaining activation-buffer capacity, and host batching mode. The algorithm is designed around engines that form a prefill-decode mixed batch, in which prefill and decode requests are processed together in one step, and extends to engines that schedule the two phases separately.

\sysname first rejects admission when fine-tuning cannot safely proceed: the backward subprocess is running, the activation buffer is full, the fine-tuning pool is empty, or, for hosts that do not form mixed batches, the pending step contains only decode work (line 5). It then computes the available latency budget $\Delta$ as the tighter of the TTFT slack and TPOT slack. The TTFT slack follows Equation~\ref{eq:budget}; the TPOT slack is $T_{\mathrm{tpot}}$ minus the predicted cost of any decode work deferred by this step. When the host forms a prefill-decode mixed batch, no decode batch is deferred and $D_{\mathrm{wait}}$ is empty (line 8-10).

\begin{algorithm}[h!]
\small
\caption{Fine-tuning admission for one step.}
\label{alg:admit}
\begin{algorithmic}[1]
\Require Host proposed inference batch $B_{\mathrm{inf}}$ for the next step; SLO estimator $E$; fine-tuning pool $Q_{\mathrm{ft}}$; Backward process state $P_{B}$; remaining buffer $S_{\max}$; host capability \textit{mixedBatch}
\Ensure Fine-tuning samples $R_{\mathrm{ft}}$ to enqueue for the next step
\State $R_{\mathrm{ft}} \gets [\,]$
\If{ $P_{B}$ is running \textbf{ or } $S_{\max} \le 0$ \textbf{ or } $Q_{\mathrm{ft}} = \emptyset$}
    \State \Return $R_{\mathrm{ft}}$
\EndIf
\If{\textbf{not} \textit{mixedBatch} \textbf{ and } $B_{\mathrm{inf}}$ has no prefill tokens}
    \State \Return $R_{\mathrm{ft}}$
    \Comment{cannot merge fine-tuning into a decode-only step}
\EndIf
\State $\Delta_{\mathrm{tt}} \gets$ TTFT budget of $B_{\mathrm{inf}}$
\Comment{Eq.~\ref{eq:budget}, with queue wait}
\State $\Delta_{\mathrm{tp}} \gets T_{\mathrm{tpot}} - E.\textsc{predict}(D_{\mathrm{wait}})$
\Comment{$D_{\mathrm{wait}}$ empty if decode shares the step}
\State $\Delta \gets \min(\Delta_{\mathrm{tt}}, \Delta_{\mathrm{tp}})$
\If{$E.\textsc{predict}(B_{\mathrm{inf}}) > \Delta$}
    \State \Return $R_{\mathrm{ft}}$
    \Comment{no slack even without fine-tuning}
\EndIf
\For{$ft \in Q_{\mathrm{ft}}$ in increasing token length}
    \State $B' \gets B_{\mathrm{inf}} \cup R_{\mathrm{ft}} \cup \{ft\}$
    \State $c_1 \gets \textsc{tokens}(R_{\mathrm{ft}}) + \textsc{tokens}(ft) > S_{\max}$
    \Comment{activation buffer full}
    \State $c_2 \gets E.\textsc{predict}(B') > \Delta$    
    \Comment{would violate an SLO}
    \If{$c_1 \lor c_2 $}
        \State \textbf{break}
    \EndIf
    \State $R_{\mathrm{ft}}.\textsc{append}(ft)$
\EndFor
\State $\textsc{Claim}(R_{\mathrm{ft}})$
\Comment{reserve buffer, remove from pool}
\State \Return $R_{\mathrm{ft}}$
\end{algorithmic}
\end{algorithm}

Given $\Delta$, \sysname first checks whether the inference-only batch already exhausts the budget (line 11-13). If not, it greedily considers fine-tuning samples in increasing token length and admits each sample only while the resulting mixed batch remains within both the latency budget and the activation-buffer capacity. The admitted samples are then claimed by reserving buffer space and removing them from the fine-tuning pool, preventing duplicate selection or buffer overcommitment under pipelined scheduling (line 14-23).

All costs are predicted by the estimator in the execution mode used by the step. The inference-only baseline is priced in the host’s native mode, which may use CUDA-graph replay, whereas any mixed batch is priced with eager-mode coefficients because activation hooks require leaving graph replay. For hosts that do not fuse prefill and decode, \sysname applies the same rule only to prefill steps and charges any delayed decode work against the TPOT budget, preserving the host’s batching semantics while exploiting prefill-phase headroom.

\vspace{1em}
\noindent\textbf{Estimator calibration.}\label{sec:calibration}
The coefficients of the latency models in Equations~\ref{eq:prefill} and~\ref{eq:decode} are initialized by an offline profiling pass before serving begins and refined online during execution. \sysname does not interfere with host engine initialization, such as CUDA-graph capture when available. Instead, it calibrates against the execution path the host will use online: after the host reports that initialization is complete, but before it begins accepting requests, \sysname issues representative synthetic batches through the same scheduling and forward path used during serving. For each observed step, \sysname records the batch composition, execution mode, and measured latency, and assigns the measurement to the corresponding coefficient set.

The profiling shapes are chosen to make the coefficients separately identifiable. \sysname sweeps total prefill length to estimate linear prefill cost, varies how a fixed prefill length is split across requests to separate quadratic self-attention from linear projection and feed-forward cost, samples a grid over decode batch size and key–value cache length, and augments inference batches with varying numbers of fine-tuning tokens to isolate activation-capture overhead. During serving, \sysname continues refining the estimator from host-reported step times, periodically refitting each execution mode so the model tracks changes in load, batching behavior, and sequence-length distribution. 

%

\vspace{1em}
\noindent\textbf{Backward preemption.}
The backward pass is compute-intensive and can interfere with inference when executed concurrently. This contention is significant during prefill, which also requires substantial compute, whereas decode is memory-bound and leaves more headroom. \sysname therefore allows backward execution to overlap with decode steps, but preempts it when prefill work arrives (\bcircled{8} in Figure~\ref{fig:pipeline}). This is implemented with a shared GPU-grant flag checked by the backward subprocess at every transformer-layer boundary. The inference process clears the flag before any forward step containing prefill tokens and restores it afterward. As a result, prefill can reclaim the GPU within one layer’s worth of backward kernels, while decode can continue to co-run with background training. 

\vspace{1em}
\noindent\textbf{Fine-tuning forward interruption.}
When no inference work is pending, \sysname may issue a forward step containing only fine-tuning samples. If an inference request arrives during such a step, \sysname sets an abort signal checked by the activation hooks at transformer-layer boundaries. The fine-tuning forward then stops, discards any partial activations, returns the interrupted samples to the fine-tuning pool, and immediately schedules inference. The interrupted samples are re-admitted later, so the cost of interruption is limited to the layers already executed.

\section{Evaluation}
In this section, we evaluate \sysname's ability to fold LoRA fine-tuning into a host inference engine under diverse conditions, harvesting the capacity inference leaves idle without violating the host's inference SLOs, guided by the following key questions:
\begin{itemize}[leftmargin=1em, itemsep=0.1em, topsep=2pt, partopsep=0pt]
	\item \textbf{End-to-end effectiveness.} Does \sysname exploit idle capacity left by gaps and low-load periods in inference workloads, converting it into fine-tuning throughput while preserving inference SLOs? How does this compare with LLMStation and with a split-pool deployment in fine-tuning throughput, average latency, and SLO compliance (Section~\ref{end-to-end})?
	
	\item \textbf{Portability across host engines.} Can \sysname's co-serving design provide benefits across different inference serving systems, rather than being tied to a single host engine? We evaluate this by integrating \sysname with vLLM, SGLang, and S-LoRA (Section~\ref{sec:portability}).

	\item \textbf{Scheduling and tunability.} How do the scheduler's per-step admission decisions shape the resulting trade-off between inference latency and fine-tuning throughput, and how does \sysname keep inference within its SLO during burst inference load (Section~\ref{sec:ablation})?
\end{itemize}

\subsection{Experimental Setup}
\noindent\textbf{Hardware.}
Our experiments run on two setups that bracket the hardware where it is likely to be deployed: a consumer-grade GPU and a datacenter-class multi-GPU server. The consumer setup is a single NVIDIA RTX~5090~(32\,GB), paired with a 16-core AMD Ryzen~9 9950X CPU (Zen~5, 4.3\,GHz) and 64\,GB of DRAM. The datacenter setup is a 4$\times$NVIDIA~A100 (40\,GB) server with a 24-core Intel Skylake Xeon CPU at 2.20\,GHz and 340\,GB of DRAM. 

\vspace{1em}
\noindent\textbf{Model.}
We evaluate on Llama 3-8B~\cite{touvron2023llamaopenefficientfoundation}, a widely deployed open-weight model with grouped-query attention, representative of the model sizes served in interactive LLM applications. The same base model is shared by inference and fine-tuning across all systems.

\vspace{1em}
\noindent\textbf{Implementations.}
Our primary implementation, \textbf{\sysname-vLLM}, realizes the host-agnostic \sysname core on vLLM version 0.21.0~\cite{kwon2023vllm}. The core is a self-contained module of roughly $4{,}000$ lines that implements the four components described in Section~\ref{sec:design}. Integrating it with vLLM requires only a compact set of in-tree hooks, confined to about a dozen host files, that expose the host's scheduling loop, forward path, and worker startup interface to \sysname. All other serving functionality is reused unchanged, including vLLM's continuous-batching scheduler, paged key--value cache, multi-LoRA forward kernels, sampler, and inference CUDA-graph capture. To demonstrate portability beyond vLLM, we also build \textbf{\sysname-SGLang} on SGLang~\cite{zheng2024sglang}, whose scheduler and execution path differ from vLLM's, and \textbf{\sysname-S-LoRA} on S-LoRA~\cite{MLSYS2024_906419cd}, a minimal multi-LoRA serving engine. The SGLang integration shows that \sysname can be adapted to another production-oriented serving stack, while the S-LoRA integration shows that the same core can be ported to a lightweight engine that provides little beyond multi-LoRA batching. Together, the three integrations show that \sysname depends only on multi-LoRA batching as a host capability, rather than on a particular serving-engine architecture.

\vspace{1em}
\noindent\textbf{Baselines.}
We compare against two baselines. \emph{LLMStation}~\cite{he2025resource} is the state-of-the-art inference–fine-tuning co-serving system and our primary point of comparison; it exploits decode-side opportunities to co-schedule fine-tuning with inference. The second baseline is a split-pool deployment, \emph{vLLM+torchtune}, which reflects a common production approach: vLLM serves inference on one GPU pool, while torchtune 0.6.1~\cite{torchtune} runs LoRA fine-tuning on a separate dedicated GPU, with no coordination between the two workloads.

\vspace{1em}
\noindent\textbf{Fine-tuning Workload.}
For all co-serving experiments, we fine-tune a dedicated LoRA adapter on the Llama 3-8B base model. The adapter uses a standard configuration with rank $r=16$, scaling factor $\alpha=32$, dropout $0.05$, and is applied to all attention projection matrices. Fine-tuning is performed on the Alpaca instruction-tuning dataset~\cite{alpaca}, a corpus of instruction--response pairs that is widely used as a LoRA fine-tuning benchmark~\cite{he2025resource, oliaro2026flexllm}. For all \sysname-enabled systems, training uses the \texttt{AdamW} optimizer with a learning rate of $1\times 10^{-3}$ and weight decay $0.01$. The backward batch size is set to 256 tokens, meaning activations are accumulated until 256 fine-tuning tokens are processed before a backward pass is triggered.

\begin{figure}[h]
  \centering
    \includegraphics[width=\columnwidth]{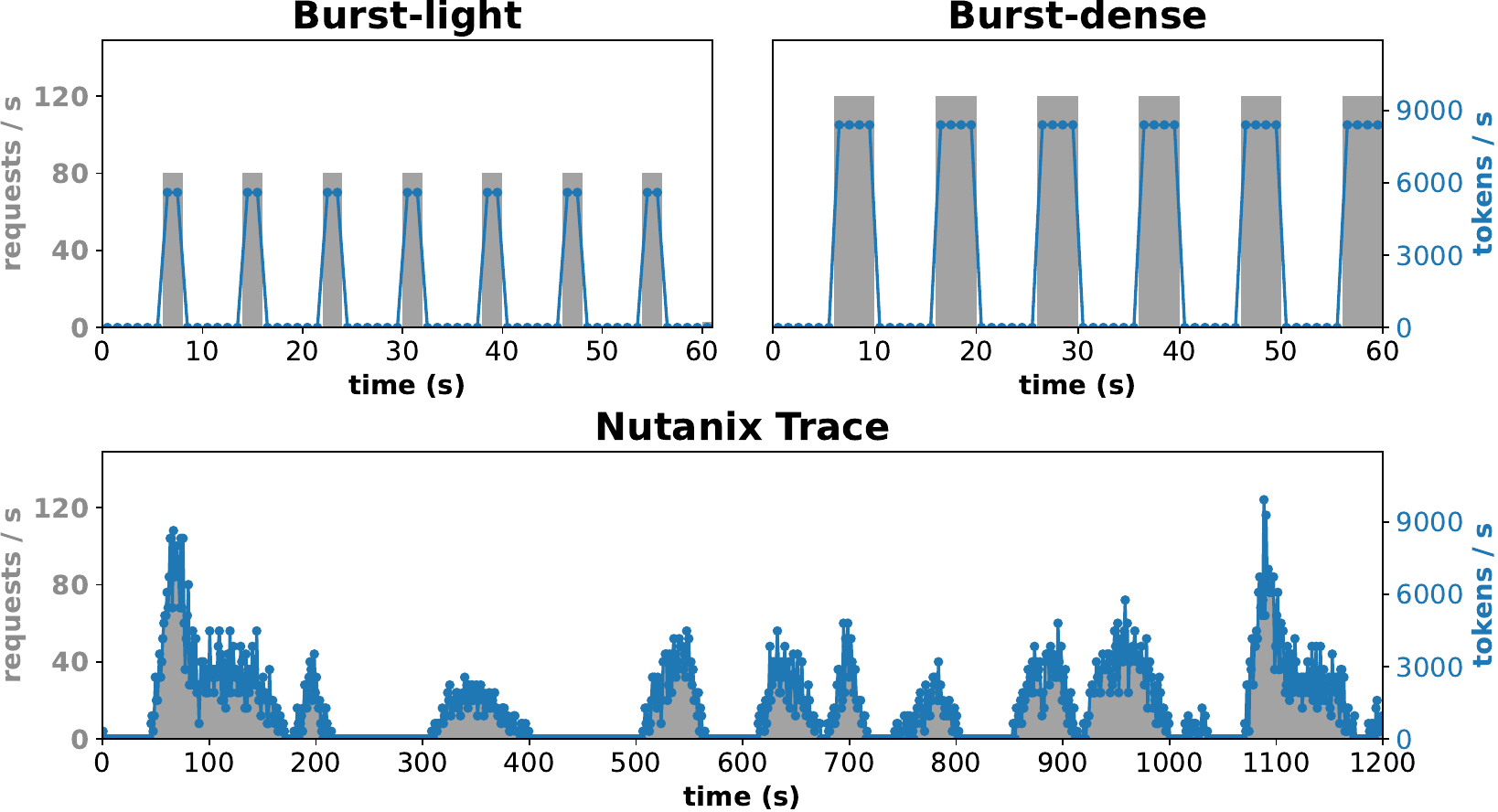}
  \caption{Inference workloads for evaluation: the synthetic burst-light and burst-dense patterns (top) and the 20-minute Nutanix production trace (bottom). In each panel the grey shaded bars report the incoming request rate and the blue line reports the resulting output-token rate.
  }
  \label{fig:workloads}
\end{figure}

\begin{figure*}[t!]
  \centering

  \includegraphics[width=\linewidth]{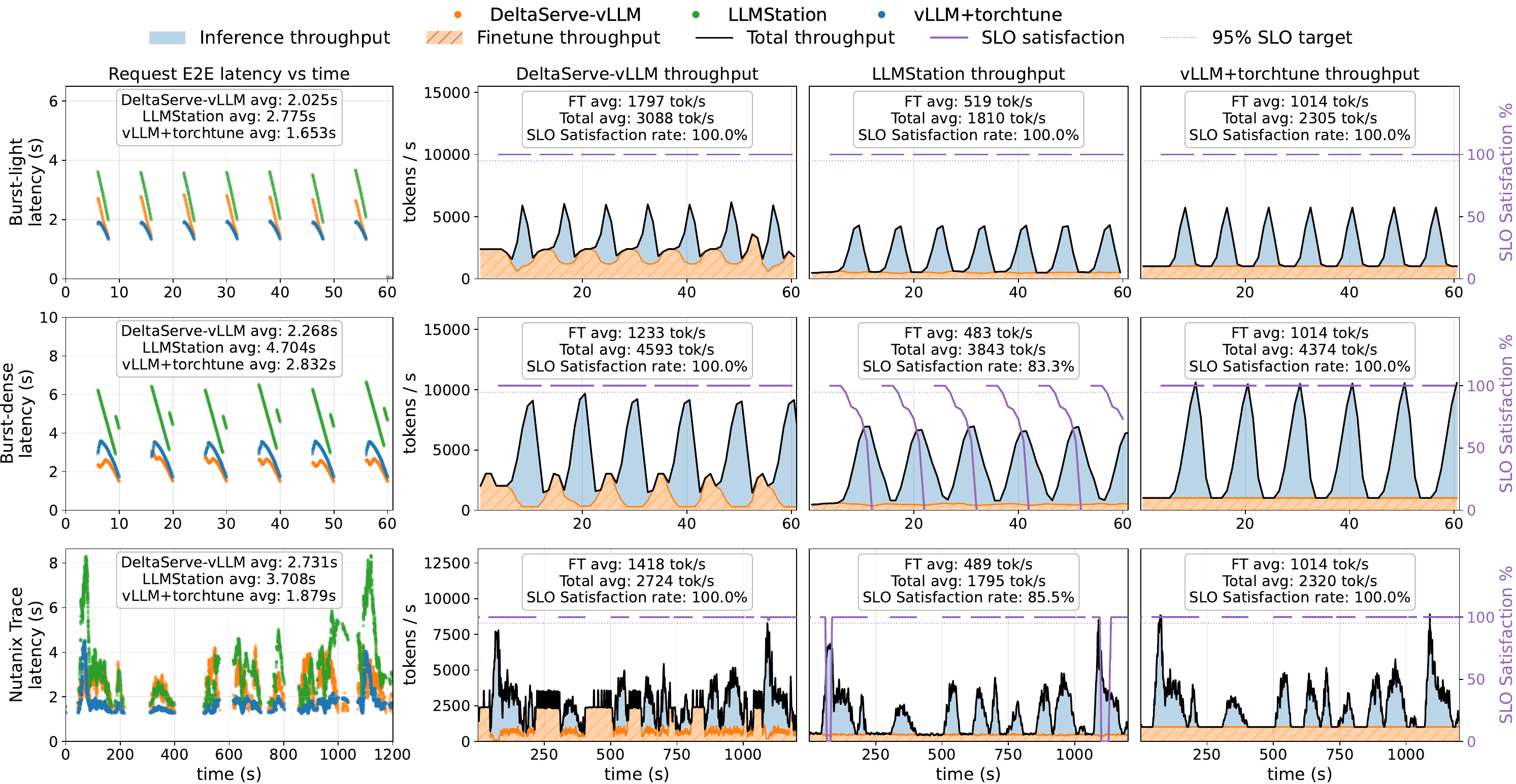}
  \caption{End-to-end comparison of \sysname-vLLM, LLMStation, and vLLM+torchtune on the 4-GPU deployment, one row per workload (top: burst-light; middle: burst-dense; bottom: Nutanix). The left column overlays per-request end-to-end latency against arrival time, with per-system averages annotated. The three right columns show each system's per-second token throughput as a stacked band: fine-tuning (orange hatched) below inference (blue), the black line tracing total throughput (3s rolling mean). The purple curve (right axis) tracks the rolling fraction of requests meeting the TTFT/TPOT SLO, against a dotted 95\% target. Each panel reports mean fine-tuning throughput, mean total throughput, and overall SLO satisfaction.}
  \label{fig:comparison-summary}
\end{figure*}

\vspace{1em}
\noindent\textbf{Inference Workloads.}
\label{inf-workload}Figure~\ref{fig:workloads} summarizes the three inference workloads used in our evaluation: two one-minute synthetic burst patterns, \emph{burst-light} and \emph{burst-dense}, and a 20-minute production trace from Nutanix. In each panel, the x-axis shows wall-clock time, the shaded grey area reports incoming request rate on the left y-axis, and the blue curve reports output-token rate on the right y-axis. The synthetic workloads differ in both burst intensity and burst duration. \emph{Burst-light} uses short 2-second peaks at 80 requests per second (RPS), leaving frequent low-load intervals and exposing substantial headroom for co-serving. \emph{Burst-dense} uses longer 4-second peaks at $120$~RPS, increasing the fraction of time spent under high load and reducing the scheduling slack available for fine-tuning. The Nutanix trace is irregular, with multi-scale bursts, spikes above $100$~RPS, and extended near-zero intervals, providing a production-like test of \sysname under realistic load variation. These profiles are used as the primary four-GPU workloads; for single-GPU experiments, we replay the same temporal patterns with request rates scaled down proportionally, preserving burst structure while matching the capacity of one GPU. We set the service-level objectives to $400$ms TTFT and $120$ms maximum TPOT on A100 systems, and to $200$ms TTFT and $100$ms maximum TPOT on RTX~5090.


\subsection{End-to-End Evaluation}
\label{end-to-end}
We first evaluate \sysname in the multi-GPU setting targeted by production deployments. 
\sysname-vLLM co-serves inference and fine-tuning on all four GPUs under analytic SLO-aware admission control. LLMStation also co-serves the two workloads, but admits fine-tuning through decode-side opportunities using a cache-based SLO estimator. The split-pool reference, vLLM+torchtune, dedicates three GPUs to vLLM inference and one GPU to torchtune fine-tuning.
Since all systems process the same arrivals, their inference token throughput is comparable within each workload; \textbf{the key differences are how much fine-tuning throughput they obtain and how that throughput affects inference latency and SLO compliance.}

Figure~\ref{fig:comparison-summary} reports the comparison, one row per workload, against the timelines of Figure~\ref{fig:workloads}. The leftmost column overlays per-request end-to-end latency for all three systems, with per-system averages boxed; the remaining three columns give each system its own throughput panel, stacking fine-tuning tokens beneath inference tokens with the rolling TTFT-SLO satisfaction overlaid against a $95\%$ guide. 

\noindent\textbf{Burst-light.}
The burst-light trace has the shortest and lowest-intensity peaks, giving co-serving systems frequent opportunities to use otherwise idle GPU capacity. All systems satisfy the SLO for every request on this workload, but \sysname-vLLM exploits the available headroom most effectively, reaching the highest fine-tuning throughput at $1797$~tok/s: $3.5\times$ that of LLMStation and $77\%$ above the split-pool baseline. This throughput comes at modest inference cost. \sysname-vLLM’s average end-to-end latency is only $22\%$ higher than the vLLM+torchtune baseline, whereas LLMStation incurs $68\%$ overhead because its decode-phase fine-tuning inflates latency during bursts. The split-pool baseline attains the lowest latency because one GPU is dedicated to fine-tuning and the remaining three vLLM GPUs are still sufficient for this light workload. However, its fine-tuning rate is fixed by the dedicated training GPU and cannot adapt to inference slack. In contrast, \sysname-vLLM’s fine-tuning band contracts near burst peaks and expands as bursts drain, showing the admission scheduler throttling fine-tuning to preserve the SLO and reopening capacity as inference pressure subsides. This throttling also explains why \sysname-vLLM’s latency approaches the split-pool baseline near burst tails: once fine-tuning admission is reduced, late-arriving requests experience less co-serving interference.

\noindent\textbf{Burst-dense.}
The burst-dense trace increases both burst intensity and duration, sustaining 120~RPS peaks for longer periods than burst-light. Under this heavier load, preserving the inference SLO becomes the primary challenge. LLMStation fails to do so, satisfying the SLO for only $83\%$ of requests, because its co-serving policy continues to consume capacity for fine-tuning when inference is already constrained. \sysname-vLLM maintains $100\%$ SLO satisfaction by giving inference priority under peak load. This also lets \sysname-vLLM outperform the split-pool baseline: vLLM+torchtune serves inference on only three GPUs, whereas \sysname-vLLM can use the full four-GPU pool when fine-tuning is curtailed, yielding $20$\% lower average end-to-end latency.

The throughput timeline explains how \sysname-vLLM still obtains substantial fine-tuning throughput under burst-dense load. During peak intervals, fine-tuning throughput drops sharply, showing that the scheduler suppresses fine-tuning when inference pressure is high. Once each burst is absorbed, the scheduler reopens fine-tuning admission. These recovery periods account for most of the $1233$~tok/s that \sysname-vLLM achieves. On average, \sysname-vLLM achieves $2.6\times$ the fine-tuning throughput of LLMStation and $21\%$ more than the split-pool baseline. 

\noindent\textbf{Nutanix.}
The Nutanix production trace is irregular and heavy-tailed, with long stretches of light load punctuated by sharp, heavy bursts. \sysname-vLLM again achieves the highest fine-tuning throughput at $1418$~tok/s, satisfying the SLO for every request: it outpaces LLMStation by $2.9\times$ and exceeds the dedicated fine-tuning GPU of the split-pool reference by $39\%$. Harvesting this fine-tuning raises its average end-to-end latency to $45\%$ above the split-pool reference as expected since \sysname is designed to harvest the gap between execution time and SLO requirement. LLMStation incurs a larger inference cost, running $97\%$ above the reference and meeting the SLO for only $85\%$ of requests.

The latency and throughput timelines show \sysname-vLLM adapting to the trace’s changing load. During heavy bursts, such as the spike around $60$s and the sustained interval between $970$ and $1100$s, the scheduler quickly throttles both fine-tuning admission and backward execution to protect inference. Requests arriving in these windows therefore see latency close to the split-pool reference, lower during the $60$s spike and comparable during the later sustained burst, while LLMStation violates the SLO. During the longer, lower-rate intervals between roughly $300$ and $1000$s, where the request rate remains below the burst-light peak, \sysname-vLLM uses the available headroom more aggressively. This increases fine-tuning throughput and occasionally raises latency above LLMStation’s within the same window, but still remains within the SLO. 

\subsection{SGLang \& S-LoRA}
\label{sec:portability}
\begin{figure}[h]
  \centering
  \includegraphics[width=\columnwidth]{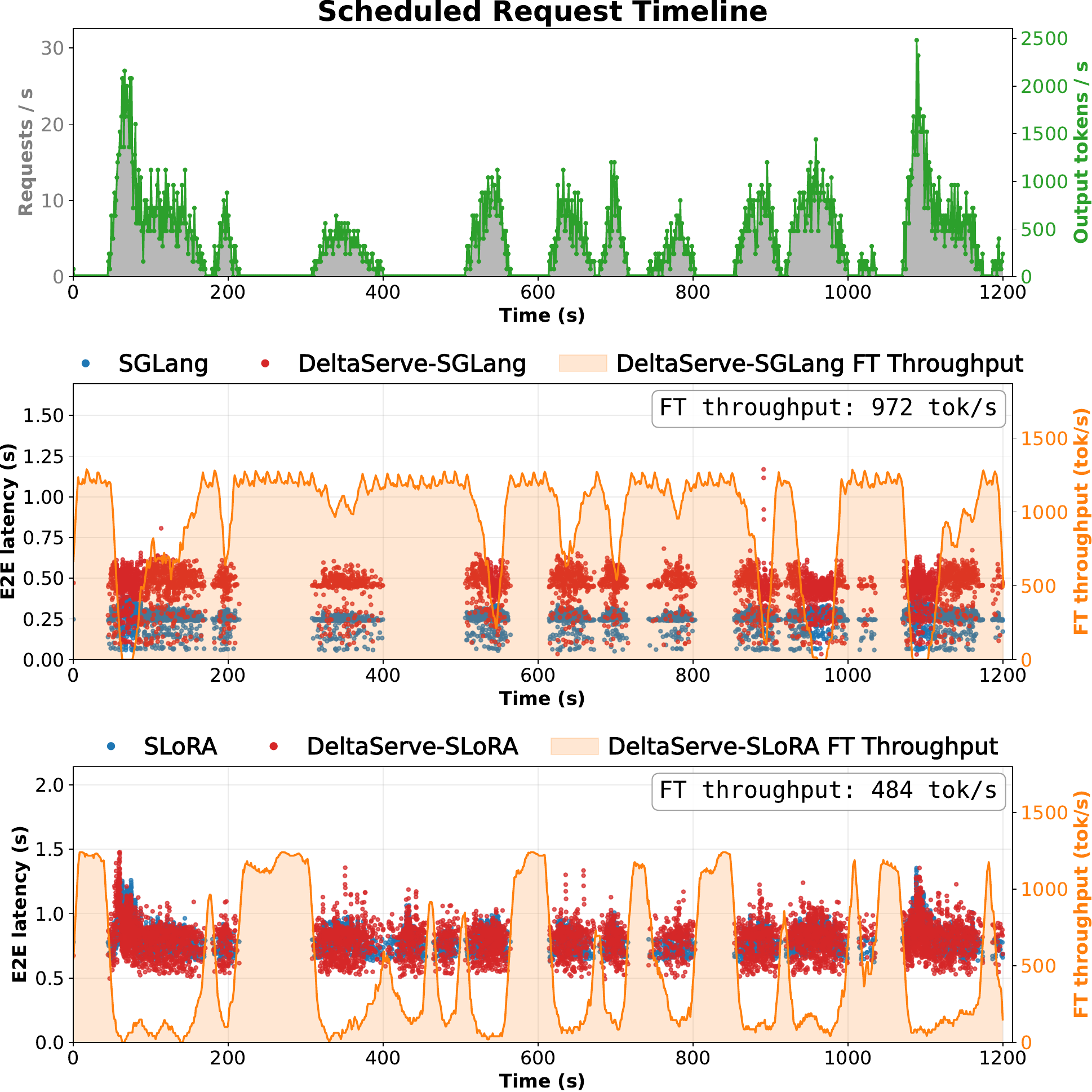}
  \caption{Portability of \sysname to SGLang and S-LoRA on a single RTX~5090, replaying the scaled-down Nutanix trace. \emph{Top:} request timeline. \emph{Middle and bottom:} per-request end-to-end latency as dots (left axis) for the unmodified host engine and its \sysname-enabled version, with fine-tuning throughput as the filled curve (right axis).}
  \label{fig:portability}
\end{figure}

We next evaluate whether \sysname’s co-serving design transfers across host engines with different serving architectures. We evaluate \sysname-SGLang and \sysname-S-LoRA on a single RTX~5090, using the scaled-down Nutanix trace from Section~\ref{inf-workload}. For each host, we compare the \sysname-enabled version against the unmodified inference-only engine. Figure~\ref{fig:portability} shows the request timeline, per-request end-to-end latency, and fine-tuning throughput over time.

SGLang represents a production-oriented inference engine with an optimized scheduling and execution path. Its runtime is designed for high-throughput LLM serving and includes optimizations that reduce inference overhead and leave more usable headroom under the same SLO. On this host, \sysname sustains $972$~tok/s of fine-tuning throughput while preserving the inference SLO for every request. The fine-tuning throughput follows the workload shape: it increases during low-load intervals, drops during request bursts, and resumes as the burst drains, mirroring the Nutanix behavior observed in the multi-GPU results of Section~\ref{end-to-end} and Figure~\ref{fig:comparison-summary}.

S-LoRA provides a complementary portability point. Unlike SGLang, it is a lightweight research serving engine centered on multi-LoRA batching, with fewer inference-side optimizations, making it close to the minimum host capability \sysname requires. On this host, \sysname still sustains $484$~tok/s of fine-tuning throughput with full SLO compliance. The lower throughput is expected: with a less optimized host inference pipeline, the same inference trace consumes more of the latency budget, leaving less headroom for fine-tuning admission. \sysname-enabled S-LoRA slightly reduces average end-to-end latency. This improvement does not come from co-serving itself, but from the CUDA-graph capture added as part of the integration. The result shows that \sysname can operate even on a minimal multi-LoRA host while preserving the host’s inference semantics.

\subsection{Ablation Study}
\label{sec:ablation}
We next evaluate \sysname in isolation to separate the effects of its individual mechanisms. Using scaled-down versions of the same traces on a single RTX~5090, we study two components without the averaging effect of multi-GPU replication. First, we measure the benefit and latency cost of admitting fine-tuning into inference-carrying batches rather than restricting it to idle steps (\emph{forward batch co-serving}).  We then evaluate how \sysname protects inference latency when a request arrives during an in-flight fine-tuning-only forward pass (\emph{fine-tuning forward interruption}).

\begin{figure}[h]
  \centering
  \includegraphics[width=\columnwidth]{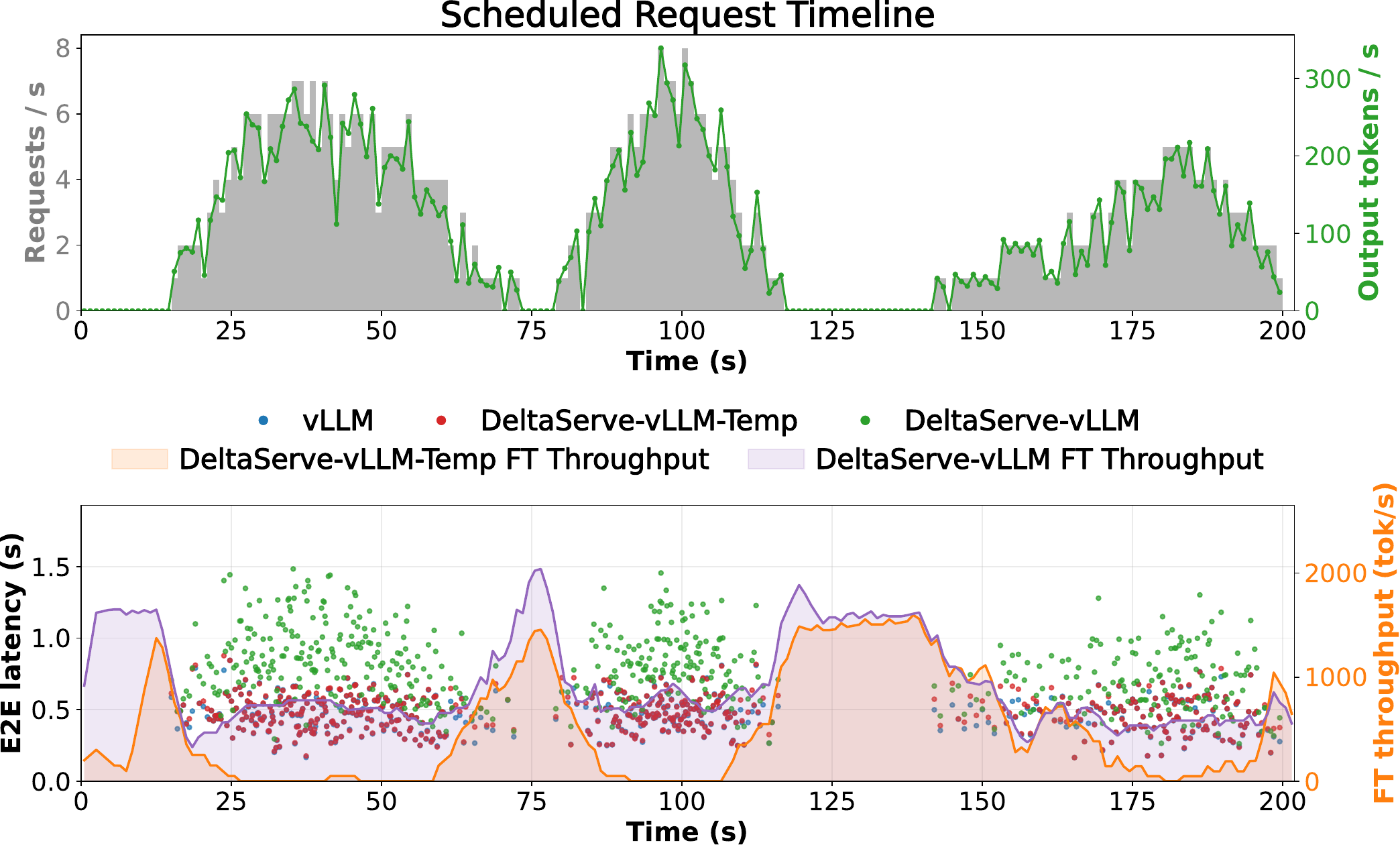}
  \caption{Forward batch co-serving on a single RTX~5090 using the $600$--$800$s window of the Nutanix trace, scaled to one RTX~5090. Top: scheduled request timeline. Bottom: per-request end-to-end latency for vLLM, \sysname-Temp, and \sysname-vLLM, with fine-tuning throughput for the two \sysname variants. \sysname-Temp restricts fine-tuning to inference-free steps.}
  \label{fig:forward-coserve}
\end{figure}

\vspace{1em}
\noindent\textbf{Forward Batch Co-serving.}
This experiment isolates the contribution of \emph{forward batch co-serving}, that is, folding fine-tuning tokens into a batch that already carries inference workload. We compare three configurations on a single RTX~5090 over the $600$--$800$s window of the Nutanix trace, with the request rate scaled down to fit the capacity of one RTX~5090. The configurations are: bare \textbf{vLLM}, which serves inference only; \textbf{\sysname-Temp}, which disables forward batch co-serving and admits fine-tuning only when the host would otherwise issue no inference work; and \textbf{\sysname-vLLM}, the full design, which also admits fine-tuning tokens into inference-carrying batches. Figure~\ref{fig:forward-coserve} reports per-request end-to-end latency and fine-tuning throughput against the same scheduled request timeline.

The orange curve in the lower panel shows that \sysname-Temp sustains substantial fine-tuning throughput even when forward batch co-serving is disabled. In this mode, \sysname exploits not only the long idle intervals in the request timeline, such as $0$--$15$s, around 75s, and around 125s, but also short gaps between inference steps within busy periods. These finer-grained opportunities appear around 40s and throughout $150$--$200$s, where \sysname-Temp continues to produce fine-tuning throughput despite nonzero inference load. Since fine-tuning-only steps never share a batch with inference, their impact on inference performance is minimal: average end-to-end latency increases by only $2\%$. Thus, when inference efficiency is prioritized, \sysname closely matches vLLM’s latency behavior while obtaining $507$~tok/s of fine-tuning throughput from otherwise unused capacity.

The few requests for which \sysname-Temp's latency exceeds bare vLLM's coincide with fine-tuning-throughput peaks, indicating that they arrived while a fine-tuning-only step was already executing and waited briefly for it to yield. The resulting delay is small because \sysname interrupts fine-tuning-only execution at fine granularity, as described in Section~\ref{sec:slo}. We evaluate the effect of this interruption mechanism separately later.

Enabling forward batch co-serving shifts \sysname-vLLM toward higher fine-tuning throughput. By allowing fine-tuning tokens to be appended to inference batches, the full system increases fine-tuning throughput from $507$ to $934$~tok/s, an $84\%$ improvement over \sysname-Temp, while still preserving the SLO. This additional throughput comes at higher inference latency because fine-tuning tokens lengthen shared forward steps. The trade-off is controlled by the SLO budget used for admission: a tighter budget would move \sysname-vLLM closer to the \sysname-Temp operating point, reducing latency overhead at the expense of fine-tuning throughput.

\vspace{1em}
\noindent\textbf{Fine-tuning forward interruption.}
This experiment isolates the fine-tuning-only step interruption mechanism in Section~\ref{sec:slo}. When an inference request arrives during a fine-tuning-only step, \sysname aborts the forward batch so inference does not wait for the full fine-tuning forward pass to finish. We compare bare \textbf{vLLM}, \textbf{\sysname-Temp} with interruption enabled, and \textbf{\sysname-No-INTR}, which disables interruption. 

\begin{figure}[h]
  \centering
  \includegraphics[width=\columnwidth]{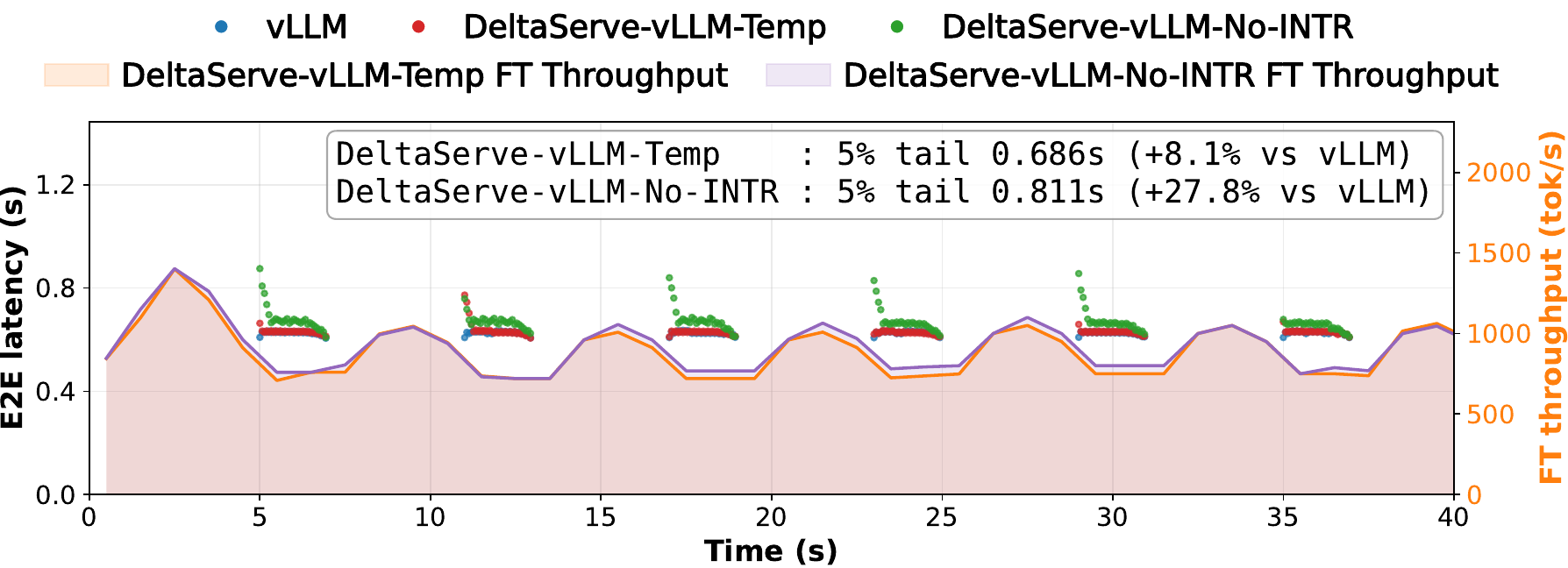}
 \caption{Effect of fine-tuning-only step interruption on one RTX~5090 over a scaled-down burst-light trace. Dots show per-request end-to-end latency; filled curves show fine-tuning throughput. Disabling interruption forces inference requests arriving during a fine-tuning-only step to wait until that step completes.}
  \label{fig:ft-interrupt}
\end{figure}
Figure~\ref{fig:ft-interrupt} shows that without interruption, latency spikes at burst onsets, where new inference requests arrive during an in-flight fine-tuning step. With interruption enabled, \sysname-Temp closely tracks bare vLLM and limits this tail amplification: average latency increases by only $0.7\%$, while the $5\%$ tail increases by $8\%$. Disabling interruption raises the $5\%$ tail overhead significantly to $27\%$. The fine-tuning throughput difference is small: \sysname-No-INTR reaches $869$~tok/s on average, only $2\%$ above \sysname-Temp. Thus, layer-boundary interruption protects inference tail latency while sacrificing little fine-tuning throughput.
\section{Conclusion}
This paper introduced \sysname, a host-agnostic co-serving design that folds LoRA fine-tuning into an existing inference engine under SLO control, driven by a CUDA-graph-aware latency model and realized on vLLM, SGLang, and S-LoRA. On a Nutanix production trace it delivers $2.9\times$ the fine-tuning throughput of LLMStation at $100\%$ inference SLO compliance versus LLMStation's $85\%$, harvesting idle GPU capacity without compromising interactive latency.



\bibliographystyle{ACM-Reference-Format}
\bibliography{references}

\end{document}